\documentclass[a4paper]{revtex4-2}
\usepackage[top=1in, bottom=1in, left=1in, right=1in]{geometry}
\usepackage{amsmath}
\usepackage{amssymb}
\usepackage{amsfonts}
\usepackage{array}
\usepackage{dcolumn}
\usepackage{color}
\usepackage[colorlinks,linkcolor=blue,citecolor=red,urlcolor=blue]{hyperref}
\usepackage[pdftex]{graphicx}
\newcommand{\be}{\begin{equation}}
\newcommand{\ee}{\end{equation}}
\newcommand{\ba}{\begin{eqnarray}}
\newcommand{\ea}{\end{eqnarray}}
\newcommand{\bea}{\begin{eqnarray*}}
\newcommand{\eea}{\end{eqnarray*}}

\begin{document}
\title{Galileon versus Quintessence: A comparative phase space analysis and late-time cosmic relevance}
\author{Mohd Shahalam$^{1,2}$\footnote{E-mail address: mohdshahamu@gmail.com}}
\affiliation{$^1$Department of Physics, Integral University, Lucknow 226026, India \\
$^2$Department of General and Theoretical Physics, L. N. Gumilyov Eurasian National University, Astana 010008, Kazakhstan
}
\begin{abstract}
\noindent We perform a comparative phase space analysis of the light mass Galileon model and standard Quintessence in the context of late--time cosmic acceleration. Focusing on a spatially flat FLRW background, we consider a cubic Galileon interaction supplemented by a scalar potential and examine three representative choices of the potential: a generalized cosh potential, a simple cosh potential, and a linear potential. By introducing suitable dimensionless variables, the cosmological field equations are reformulated as an autonomous dynamical system, allowing a systematic investigation of the stationary points and their stability properties. For the light mass Galileon scenario, we find that although the phase space admits scalar field dominated solutions, all critical points are of saddle type for the potentials considered. In particular, no stable late-time accelerating attractor emerges, even in the presence of de-Sitter like configurations. In contrast, the Quintessence limit admits stable de-Sitter attractors for cosh potentials, providing a viable description of the observed late--time acceleration. Our results highlight a key qualitative distinction between Galileon and Quintessence cosmologies and indicate that, within the light mass Galileon framework, the higher-order Galileon interactions may be required to realize a stable accelerating Universe.
\end{abstract}

\date{\today}

\maketitle

\section{Introduction}
\label{intro}
\noindent Galileon field cosmology constitutes a class of modified gravities proposed to account for the late-time accelerated expansion without invoking a cosmological constant. The framework is motivated by the decoupling limit of the Dvali Gabadadze Porrati (DGP) braneworld model and is characterized by the presence of a scalar field, known as the Galileon, whose dynamics are governed by higher derivative interactions while yielding second order equations of motion. This distinctive feature guarantees the absence of Ostrogradsky instabilities, which typically plague higher derivative theories. The Galileon theory is constructed to be invariant under the Galilean shift symmetry in Minkowski spacetime, $\phi(x) \rightarrow \phi(x) + b_{\mu} x^{\mu} + c$, where $b_{\mu}$ and $c$ are constant four vector and scalar parameters, respectively. This symmetry severely constrains the allowed interaction terms, resulting in a finite number of Lagrangian densities in flat spacetime that lead to second order field equations. When extended to cosmology, the Galileon field couples to gravity and matter, giving rise to modified Friedmann equations and scalar field dynamics capable of driving cosmic acceleration. The covariant generalization of the Galileon action ensures consistency in curved spacetime, leading to the formulation commonly referred to as Covariant Galileon Cosmology \cite{c1,c2,c3,c4,c5,c6}. In four dimensional flat spacetime, Nicolis \textit{et al.} identified five distinct Galileon Lagrangian densities ${\cal L}_i$ ($i=1,\ldots,5$) \cite{galileon}. These include the linear potential term ${\cal L}_1$, the standard kinetic term ${\cal L}_2$, and the cubic Galileon (Vainshtein) interaction ${\cal L}_3 = (\partial_{\mu}\phi)^2 \Box \phi$, which emerges naturally in the decoupling limit of the DGP \cite{dgp}. Higher order nonlinear derivative self interactions involving four and five powers of the scalar field are represented by ${\cal L}_4$ and ${\cal L}_5$, respectively. It has been shown that the existence of stable de~Sitter solutions requires the presence of at least one higher order Galileon interaction \cite{galileon3}. The cosmological dynamics of Galileon and Horndeski theories have been extensively explored in the literature. Kase and Tsujikawa \cite{k1} performed a systematic analysis of scalar tensor cosmologies within the generalized Horndeski framework, deriving stability conditions and identifying viable scenarios for late-time acceleration. Kobayashi’s comprehensive review \cite{T1} further summarized developments in Horndeski and beyond Horndeski theories, emphasizing their rich phenomenology, screening mechanisms such as the Vainshtein effect. De Felice and Tsujikawa \cite{DeFelice:2010nf} introduced the covariant Galileon model and demonstrated the emergence of self accelerating solutions without a cosmological constant. Subsequent studies examined Galileon gravity as an alternative to dark energy \cite{Appleby:2011aa}, tested non-minimally coupled Galileon models against observational data \cite{Jamil:2013nca, Linder:2012zz}, and analyzed generalized Galileon interactions and their observational viability \cite{DeFelice:2011th}. Additional works investigated de-Sitter solutions \cite{Burrage:2011bt}, matter perturbations \cite{DeFelice:2010as}, inflationary applications \cite{Burrage:2010cu}, and dynamical system approaches to Galileon cosmology \cite{Leon:2012mt}. Collectively, these studies established both the theoretical consistency and observational relevance of Galileon modified gravity.

In the present work, we focus exclusively on the cubic Galileon term ${\cal L}_3$ supplemented by a general scalar potential, commonly referred to as the light mass Galileon (LMG) scenario \cite{LMG1,LMG2,LMG3,LMG4}. Our analysis differs from most existing studies, which primarily investigate covariant Galileon models or higher order Horndeski Lagrangians involving ${\cal L}_4$ and ${\cal L}_5$. By restricting attention to the lowest order Galileon interaction combined with a potential, we aim to capture late-time cosmic acceleration within the simplest possible Galileon framework.
We consider both tracker (e.g., hyperbolic cosine) and thawing (e.g., linear potential) types of scalar field potentials to investigate the cosmological dynamics in the LMG scenario. The cosh potential explicitly breaks the Galileon shift symmetry, whereas the linear potential preserves it. The choice of cosh-type and linear potentials is motivated by the following considerations:
\begin{itemize}
\item Cosh potentials are well studied in the dark energy literature, as they interpolate between exponential behavior at large field values and power--law behavior near the origin, allowing both scaling and accelerated solutions.
\item These potentials have been previously confronted with observations in the Quintessence context \cite{LMG3,mnras,Yang:2018xah}.
\item Our aim is not to exhaust all possible potentials, but to assess whether the phenomenologically successful Quintessence potentials can provide the stable de-Sitter solution within the LMG framework.
\end{itemize}

Scalar field models provide a natural and well studied framework for dynamical dark energy. Broadly, these models are classified into two categories: canonical and non-canonical scalar field. In this work, we focus on a particular class of canonical scalar field models, namely quintessence, which is described by a single, minimally coupled scalar field with a standard kinetic term of positive sign. In contrast, phantom scalar field model are characterized by a negative kinetic term. Canonical scalar field models are often regarded as more transparent and predictive than their non-canonical counterparts, since much of their physical behavior is encoded in the scalar field potential. Owing to their ability to account for different phases of cosmic evolution, such models have attracted significant attention in cosmology \cite{z1,z2,z3,z4,z5,z6,z7,z8}. Depending on the functional form of $V(\phi)$, quintessence models can exhibit tracking or thawing behavior, thereby alleviating fine-tuning and coincidence problems associated with dark energy. Moreover, the dynamical nature of quintessence allows the deviations from $\Lambda$CDM at both background and perturbative levels, making it testable against high-precision cosmological observations. Consequently, confronting well motivated quintessence potentials with current data provides a crucial avenue for probing the physics of dark energy and assessing the viability of scalar field descriptions of cosmic acceleration. Dynamical systems analysis provides a powerful mathematical tool for investigating the evolution of cosmological models. By reformulating the cosmological field equations as an autonomous system of first order differential equations, one can analyze the qualitative behavior of solutions without requiring exact analytic expressions. The resulting phase space represents all possible states of the Universe  with fixed points corresponding to asymptotic cosmological regimes such as matter, radiation or dark energy dominated epochs. Linear perturbation techniques are employed to assess the stability of these critical points, classifying them as stable attractors, unstable repellors or saddle points. Several studies have applied dynamical systems methods to scalar field and modified gravity cosmologies. León \textit{et al.} \cite{g1} investigated the early-time dynamics of quintom cosmology, while Paliathanasis \textit{et al.} \cite{g2} revisited dark matter phantom interactions using compactified variables. More recently, León \textit{et al.} \cite{g3} extended this methodology to phantom field interactions, and Paliathanasis \cite{g4} studied the dynamical structure of $f(Q)$ gravity. Further analyses explored chiral phantom models \cite{g5}, and comprehensive treatments of dynamical systems theory can be found in standard text \cite{l1}.

The paper is organized as follows. In Section~\ref{sec:EOM}, we derive the equations of motion for the LMG model in a homogeneous, isotropic and spatially flat Friedmann--Lemaître--Robertson--Walker (FLRW) universe and construct the corresponding autonomous system. Section~\ref{sec:ps} is devoted to the phase space formulation. In Sections~\ref{sec:d1} and \ref{sec:d0}, we identify the critical points, analyze their stability properties, and discuss their cosmological significance. The cosmological implications, results, and discussion are presented in Section~\ref{sec:CID}, while our conclusions are summarized in Section~\ref{sec:conc}.
\section{Basics Equations and Autonomous system}
\label{sec:EOM}
\noindent The LMG model extends standard Galileon cosmology by introducing a small mass term through an explicit potential. This modification enables the scalar field to behave as a dynamical dark energy component at late times, while remaining consistent with local gravity constraints via the Vainshtein screening mechanism. Although the inclusion of a mass term softly breaks the exact Galilean shift symmetry of the original theory, the breaking is sufficiently mild to preserve key theoretical properties such as second order field equations and the absence of ghost instabilities. From a physical perspective, the LMG framework allows the scalar field to undergo slow-roll evolution on cosmological scales, thereby contributing to the accelerated expansion of the Universe. We consider the Galileon action truncated at the cubic interaction level and supplemented by a scalar potential $V(\phi)$ \cite{LMG2,LMG3}.
\begin{equation}
S=\int d^4x\sqrt{-g}\left[\frac{M_{\rm pl}^2}{2}R
-\frac{1}{2}(\nabla\phi)^2\left(1+\frac{\delta}{M^3}\Box\phi\right)
- V(\phi)\right]+\mathcal{S}_m,
\label{eq:action}
\end{equation}
where $M_{\rm pl}^2=(8\pi G)^{-1}$ is the reduced Planck mass, $\delta$ is a dimensionless coupling constant, and $\mathcal{S}_m$ denotes the matter action. Throughout this work, we set $M=M_{\rm pl}$, with $M$ being a mass scale of dimension one. The action \eqref{eq:action} contains the canonical kinetic term, the cubic Galileon interaction, and a light mass contribution encoded in the potential. Varying the action with respect to the metric $g_{\mu\nu}$ and the scalar field $\phi$ in a homogeneous, isotropic, and spatially flat FLRW background yields the following field equations:
\begin{align}
3M_{\rm pl}^2H^2 &=
\rho_m+\frac{\dot{\phi}^2}{2}\left(1-6\frac{\delta}{M_{\rm pl}^{3}}H\dot{\phi}\right)+V(\phi),
\label{eq:H}\\
M_{\rm pl}^2\left(2\dot H+3H^2\right)&=
-\frac{\dot{\phi}^2}{2}\left(1+2\frac{\delta}{M_{\rm pl}^{3}}\ddot{\phi}\right)+V(\phi),
\end{align}
and
\begin{align}
\ddot{\phi}+3H\dot{\phi}
-3\frac{\delta}{M_{\rm pl}^{3}}\dot{\phi}
\left(3H^2\dot{\phi}+\dot{H}\dot{\phi}+2H\ddot{\phi}\right)
+V_{,\phi}=0,
\end{align}
where $\rho_m$ denotes the energy density of matter. The matter sector obeys the standard conservation equation,
\begin{equation}
\dot{\rho}_m+3H\rho_m=0.
\end{equation}
To perform the dynamical system analysis, we introduce the following dimensionless variables:
\begin{align}
x&=\frac{\dot{\phi}}{\sqrt{6}HM_{\rm pl}}, \qquad
y=\frac{\sqrt{V}}{\sqrt{3}HM_{\rm pl}},\\
\epsilon&=-6\frac{\delta}{M_{\rm pl}^3}H\dot{\phi}, \qquad
\lambda=-M_{\rm pl}\frac{V_{,\phi}}{V}.
\end{align}
Using the number of $e$-folds $N=\ln a$ as the time variable, such that $\frac{d}{dt}=H\frac{d}{dN}$, the cosmological equations can be cast into the autonomous system
\begin{align}
\frac{dx}{dN}&=x\left(\frac{\ddot{\phi}}{H\dot{\phi}}-\frac{\dot H}{H^2}\right), \nonumber\\
\frac{dy}{dN}&=-y\left(\sqrt{\frac{3}{2}}\lambda x+\frac{\dot H}{H^2}\right), \nonumber\\
\frac{d\epsilon}{dN}&=\epsilon\left(\frac{\ddot{\phi}}{H\dot{\phi}}+\frac{\dot H}{H^2}\right), \nonumber\\
\frac{d\lambda}{dN}&=\sqrt{6}x\lambda^2(1-\Gamma),
\label{eq:auto}
\end{align}
where $\Gamma = \frac{V(\phi)V''(\phi)}{V'(\phi)^2}$. The auxiliary quantities appearing above are given by
\begin{align}
\frac{\dot H}{H^2}&=
\frac{2(1+\epsilon)(-3+3y^2)-3x^2(2+4\epsilon+\epsilon^2)
+\sqrt{6}x\epsilon y^2\lambda}
{4+4\epsilon+x^2\epsilon^2},
\\
\frac{\ddot{\phi}}{H\dot{\phi}}&=
\frac{3x^3\epsilon-x\left[12+\epsilon(3+3y^2)\right]
+2\sqrt{6}y^2\lambda}
{x\left(4+4\epsilon+x^2\epsilon^2\right)}.
\end{align}
The effective equation of state (EoS) parameter and the EoS of the Galileon field are defined as
\begin{align}
w_{\rm eff}&=-1-\frac{2\dot H}{3H^2},\\
w_{\phi}&=\frac{w_{\rm eff}-w_m\Omega_m}{1-\Omega_m},
\end{align}
where, $w_m=0$ corresponds to pressureless dust. The matter density parameter is expressed as
\begin{equation}
\Omega_m=1-\Omega_{\phi}=1-x^2(1+\epsilon)-y^2.
\end{equation}
In the following section, we present a detailed phase space analysis of the LMG $(\delta \neq 0)$ and Quintessence $(\delta = 0)$ with three different potentials.
\section{PHASE SPACE ANALYSIS: Stationary points and their stability}
\label{sec:ps}
\noindent Phase space analysis is a powerful method for studying the qualitative dynamics of cosmological models by reformulating the field equations as an autonomous system of first order differential equations. The resulting phase space enables a global understanding of cosmic evolution without requiring exact solutions. Critical points in the phase space correspond to important cosmological epochs such as radiation, matter and dark energy dominated eras.
\begin{table}[tbp]
\caption{The table presents the stationary points, corresponding eigenvalues, and relevant cosmological parameters for the LMG model (Eq. (\ref{eq:action}); $\delta \ne 0$) with potential (\ref{eq:Vm1}). In this table, we define $\gamma_1=\sqrt{p^8 \alpha^8(24-7 p^2 \alpha^2)}$.}
\begin{center}
\begingroup
\setlength{\tabcolsep}{6pt} 
\renewcommand{\arraystretch}{1.5} 
\begin{tabular}{c c c c c c c c c}
\hline
Point  & x  & y & $\epsilon$ & $\lambda$  & Eigenvalues & Stability  & $\Omega_{\phi}$  & $w_{eff}$\\
\hline
$A_1$ & $\pm 1$ & 0 & 0 & $\pm p \alpha$ & $
\begin{array}{l}
\mu_1 = 3 \\
\mu_2 = -6 \\
\mu_3 = \sqrt{6} \alpha \\
\mu_4 = \frac{6-\sqrt{6} p \alpha}{2} \\
\end{array}
$ & Saddle & 1 & 1\\\\
$A_2$ & $\pm \sqrt{\frac{3}{2 p^2 \alpha^2}}$ & $\pm \sqrt{\frac{3}{2 p^2 \alpha^2}}$ & 0 & $\pm p \alpha$ & $
\begin{array}{l}
\mu_1 = -3 \\
\mu_2 = 3/p \\
\mu_3 =-\frac{3(p^5 \alpha^5+\gamma_1)}{4 p^5 \alpha^5} \\
\mu_4 = -\frac{3(p^5 \alpha^5-\gamma_1)}{4 p^5 \alpha^5} \\
\end{array}
$ & Saddle & $\frac{3}{p^2 \alpha^2}$ & 0\\\\
$A_3$ & $\pm \frac{p \alpha}{\sqrt{6}}$ & $\pm \sqrt{1-\frac{p^2 \alpha^2}{6}}$ & 0 & $\pm p \alpha$ & $
\begin{array}{l}
\mu_1 = -p^2 \alpha^2 \\
\mu_2 = p \alpha^2 \\
\mu_3 =p^2 \alpha^2-3 \\
\mu_4 = \frac{p^2 \alpha^2-6}{2} \\
\end{array}
$ & Saddle & 1 & $\frac{p^2 \alpha^2}{3}-1$ \\
\hline
\end{tabular}
\endgroup
\label{tab1}
\end{center}
\end{table}
\begin{table}[tbp]
\caption{The stationary points, eigenvalues, and cosmological parameters for the LMG (Eq. (\ref{eq:action}); $\delta \ne 0$) with potential (\ref{eq:Vm2}) are summarized in the table, where $\gamma_2=\sqrt{\beta^8(24-7 \beta^2)}$.}
\begin{center}
\begingroup
\setlength{\tabcolsep}{6pt} 
\renewcommand{\arraystretch}{1.5} 
\begin{tabular}{c c c c c c c c c}
\hline
Point  & x  & y & $\epsilon$ & $\lambda$  & Eigenvalues & Stability  & $\Omega_{\phi}$  & $w_{eff}$\\
\hline
$B_1$ & $\pm 1$ & 0 & 0 & $\pm \beta$ & $
\begin{array}{l}
\mu_1 = 3 \\
\mu_2 = -6 \\
\mu_3 = \sqrt{24} \beta \\
\mu_4 = \frac{6-\sqrt{6} \beta}{2} \\
\end{array}
$ & Saddle & 1 & 1\\\\
$B_2$ & $\pm \sqrt{\frac{3}{2 \beta^2}}$ & $\pm \sqrt{\frac{3}{2 \beta^2}}$ & 0 & $\pm \beta$ & $
\begin{array}{l}
\mu_1 = -3 \\
\mu_2 = 6 \\
\mu_3 =-\frac{3(\beta^5+\gamma_2)}{4 \beta^5} \\
\mu_4 = -\frac{3(\beta^5-\gamma_2)}{4 \beta^5} \\
\end{array}
$ & Saddle & $\frac{3}{\beta^2}$ & 0\\\\
$B_3$ & $\pm \frac{\beta}{\sqrt{6}}$ & $\pm \sqrt{1-\frac{\beta^2}{6}}$ & 0 & $\pm \beta$ & $
\begin{array}{l}
\mu_1 = -\beta^2 \\
\mu_2 = 2 \beta^2 \\
\mu_3 =\beta^2-3 \\
\mu_4 = \frac{\beta^2-6}{2} \\
\end{array}
$ & Saddle & 1 & $\frac{\beta^2}{3}-1$\\
\hline
\end{tabular}
\endgroup
\label{tab2}
\end{center}
\end{table}
\begin{table}[tbp]
\caption{The table lists the stationary points along with their associated eigenvalues and cosmological parameters for the LMG model (Eq. (\ref{eq:action}); $\delta \ne 0$) considering the linear potential (\ref{eq:Vm3}).}
\begin{center}
\begingroup
\setlength{\tabcolsep}{6pt} 
\renewcommand{\arraystretch}{1.5} 
\begin{tabular}{c c c c c c c c c}
\hline
Point  & x  & y & $\epsilon$ & $\lambda$  & Eigenvalues & Stability  & $\Omega_{\phi}$  & $w_{eff}$\\
\hline
$C_1$ & $\pm 1$ & 0 & 0 & 0 & $
\begin{array}{l}
\mu_1 = 0 \\
\mu_2 = 3 \\
\mu_3 = 3 \\
\mu_4 = -6 \\
\end{array}
$ & Saddle & 1 & 1\\\\
$C_2$ & 0 & $\pm 1$ & $\epsilon=-2$ & 0 & $
\begin{array}{l}
\mu_1 = 0 \\
\mu_2 = 0 \\
\mu_3 =-3 \\
\mu_4 = -3\\
\end{array}
$ & Saddle & 1 & $-1$\\
\hline
\end{tabular}
\endgroup
\label{tab3}
\end{center}
\end{table}
The critical points of the autonomous system (\ref{eq:auto}) are obtained by simultaneously solving
\begin{equation}
\frac{dx}{dN}=\frac{dy}{dN}=\frac{d\epsilon}{dN}=\frac{d\lambda}{dN}=0.
\end{equation}
To examine the stability of each critical point, we linearize the autonomous system in the vicinity of the fixed points and evaluate the eigenvalues of the associated Jacobian matrix. The nature of a critical point is classified according to the following criteria:
\begin{itemize}
\item \emph{Stable} (attractor), if all eigenvalues possess negative real parts.
\item \emph{Unstable} (repellor), if all eigenvalues have positive real parts.
\item \emph{Saddle}, if eigenvalues with both positive and negative real parts are present.
\end{itemize}
These conditions enable the identification of physically viable late-time attractor solutions that can account for the accelerated expansion of the Universe. In the present analysis, we consider two distinct regimes characterized by the Galileon coupling parameter $\delta$. Specifically, we investigate the case $\delta=0$, corresponding to the quintessence scenario, and the case $\delta\neq0$, which describes the LMG model. In both scenarios, we analyze three different choices of the scalar potential. The corresponding dynamical behavior and stability properties are discussed in Subsections~\ref{sec:d1} and \ref{sec:d0}, respectively.
\subsection{Light Mass Galileon: $\delta \neq 0$}
\label{sec:d1}
\noindent Galileon theories generalize standard scalar field models by incorporating higher-derivative self interactions that lead to second order equations of motion, thereby avoiding Ostrogradsky instabilities. These theories naturally arise in modified gravity scenarios and braneworld constructions. In LMG framework, the scalar field is endowed with a potential, so that its cosmological dynamics are governed by a nontrivial interplay between the potential energy and derivative self nteractions on cosmological scales. The Galileon correction term appearing in the Friedmann equation~(\ref{eq:H}) is proportional to $3\delta H\dot{\phi}^3/M_{\rm pl}^3$. When $\delta\neq 0$, the Galileon correction term becomes dynamically significant and modifies the standard quintessence evolution.
\subsubsection{Model 1: $V(\phi)=V_0\left[\cosh\left(\frac{\alpha \phi}{M_{pl}}\right)-1\right]^p$}
\label{sec:m1}
\noindent We consider a Galileon scalar field endowed with a $cosh$ type potential. This setup is purely phenomenological and is introduced to allow greater flexibility and tractability in the analysis. It should be emphasized that such a potential explicitly breaks the Galileon shift symmetry,  however, this feature is common in scalar field models, where the choice of potential is typically phenomenological rather than symmetry driven. In this context, we consider the following potential proposed by Sahni and Wang \cite{sahni, mnras}:
\begin{equation}
V(\phi)=V_0\left[\cosh\left(\frac{\alpha \phi}{M_{pl}}\right)-1\right]^p;
\qquad 0<p<\frac{1}{2},
\label{eq:Vm1}
\end{equation}
where $V_0$, $\alpha$ and $p$ are positive constants. To obtain cosmic acceleration the range of $p$ should be $0<p<\frac{1}{2}$ \cite{sahni, mnras}. The potential (\ref{eq:Vm1}) admits the following asymptotic forms:
\begin{equation}
V(\phi)=\frac{V_0}{2^p}\exp\left(\frac{p \alpha \phi}{M_{pl}}\right),
\qquad \frac{p \alpha |\phi|}{M_{pl}}\gg1,
\label{exppot}
\end{equation}
and
\begin{equation}
V(\phi)=\frac{V_0}{2^p}
\left(\frac{\alpha \phi}{M_{pl}}\right)^{2p},
\qquad \frac{p \alpha |\phi|}{M_{pl}}\ll1,
\label{ppot}
\end{equation}
The power law behavior of the potential (\ref{eq:Vm1}) leads to oscillations of the scalar field as it approaches $\phi=0$. For this potential, we have
\ba
\Gamma &=& \frac{V(\phi)V''(\phi)}{V'(\phi)^2}= 1-\frac{1}{2p}+\frac{p \alpha^2}{2 \lambda^2}.
\label{eq:gamma1}
\ea
The autonomous system (\ref{eq:auto}) together with Eq. (\ref{eq:gamma1}) can be expressed in terms of phase space variables. The critical points of the resulting dynamical system are obtained by setting the left-hand side of the autonomous equation~(\ref{eq:auto}) to zero. The corresponding critical points, along with their associated eigenvalues and cosmological parameters are summarized in Table~\ref{tab1}. Below is a detailed physical and dynamical interpretation of each critical point.

$A_1$: This point represents a pure kinetic energy dominated scalar field. The potential contribution vanishes ($y=0$), and the Universe behaves like a stiff fluid. Such a solution is typical of early-time cosmology. The eigenvalues contain both positive and negative real parts, which makes the point a saddle node. Hence, it cannot describe the late-time Universe. 

$A_2$: This point corresponds to a scaling behavior, where the scalar field energy density scales like pressureless matter. The effective equation of state mimics dust, making it relevant for the matter dominated epoch. The eigenvalues include both positive and negative components, ensuring that the point is saddle.

$A_3$: This point corresponds to a scalar field dominated Universe, where both kinetic and potential energies contribute. Depending on the value of $p^2 \alpha^2$ the model can exhibit: accelerated expansion for $p^2 \alpha^2 < 2$, otherwise decelerated expansion. However, the presence of eigenvalues with opposite signs makes this point a saddle, preventing it from being a late-time attractor.
\subsubsection{Model 2: $V(\phi)=V_0\,\cosh\!\left(\frac{\beta \phi}{M_{pl}}\right)$}
\label{sec:m2}
\noindent We now consider a scalar field model characterized by the potential \cite{Yang:2018xah}.
\begin{equation}
V(\phi)=V_0\,\cosh\!\left(\frac{\beta \phi}{M_{pl}}\right),
\label{eq:Vm2}
\end{equation}
where $\beta$ represents a real dimensionless parameter, and quantifies the deviation of the potential from a constant energy $V(\phi)=V_0$. The potential (\ref{eq:Vm2}) can be written equivalently as a sum of two exponential terms,
\begin{equation}
V(\phi)=\frac{V_0}{2}\left[
\exp\!\left( \frac{\beta \phi}{M_{pl}}\right)
+\exp\!\left(- \frac{\beta \phi}{M_{pl}}\right)
\right],
\end{equation}
and, the asymptotic behavior is given by
\begin{equation}
V(\phi)\simeq \frac{V_0}{2}
\exp\!\left(\frac{\beta\phi}{M_{pl}}\right),
\qquad \frac{\beta|\phi|}{M_{pl}}\gg1,
\label{pot1exp}
\end{equation}
and
\begin{equation}
V(\phi)\simeq V_0\left[1+\frac{1}{2}
\left(\frac{\beta\phi}{M_{pl}}\right)^2\right],
\qquad \frac{\beta|\phi|}{M_{pl}}\ll1.
\label{pot1sq}
\end{equation}
For this potential, we have
\ba
\Gamma &=& \frac{\beta^2}{\lambda^2}.
\label{eq:gamma2}
\ea
In Ref. \cite{Yang:2018xah}, we examined the qualitative cosmological behavior of this potential under observational constraints. In the present study, we focus on identifying the stationary points of the model and analyzing their stability properties. The critical points of the autonomous system (\ref{eq:auto}) with equation (\ref{eq:gamma2}) are obtained by setting the left-hand side equal to zero. These points, together with their corresponding eigenvalues and cosmological parameters are summarized in Table \ref{tab2}. A detailed physical and dynamical interpretation of each critical point is presented below.

$B_1$: This point has the same physical interpretation as $A_1$. The Universe is dominated by the kinetic energy of the scalar field and behaves as a stiff fluid. The mixed sign eigenvalues again classify it as a saddle.

$B_2$: This solution describes a matter dominated era, where the scalar field effectively behaves as cold dark matter. The point shows saddle nature.

$B_3$: This point corresponds to a scalar field dominated cosmology with the possibility of accelerated expansion for $\beta^2 < 2$. Despite this, the mixed eigenvalue structure renders it a saddle node.
\subsubsection{Model 3: $V(\phi)=V_0 \left(\frac{\phi}{M_{pl}}\right)$}
\label{sec:m3}
\noindent In cosmology, a linear potential refers to a scalar field potential that depends linearly on the field,
\begin{equation}
V(\phi)=V_0 \left(\frac{\phi}{M_{pl}}\right),
\label{eq:Vm3}
\end{equation}
Such potentials arise naturally in effective field theory descriptions and can also emerge as leading order approximations of more complicated potentials over a restricted field range. Linear potentials do not break Galileon shift symmetry, and are often employed as phenomenological models due to their simplicity and analytical tractability. In the context of dark energy and inflation, linear potentials allow the scalar field to evolve slowly over cosmological scales, leading to accelerated expansion when the potential energy dominates over the kinetic term. Unlike exponential or power law potentials, linear potentials do not typically admit exact scaling solutions, however, they can support quasi de Sitter evolution and transient acceleration. These models are particularly useful for studying departures from $\Lambda$CDM and for exploring the sensitivity of cosmic dynamics to mild symmetry breaking in scalar field theories. In this case,
\ba
\Gamma &=& 0.
\label{eq:gamma3}
\ea
By equating the left-hand side of the autonomous system (\ref{eq:auto}) with equation (\ref{eq:gamma3}) to zero, we determine the critical points of the model. The resulting stationary solutions along with their eigenvalues and cosmological parameters are listed in Table \ref{tab3}, followed by a detailed discussion of their physical and dynamical properties.

$C_1$: This solution again represents a stiff fluid dominated universe driven entirely by the scalar field’s kinetic energy. The presence of zero and positive eigenvalues indicates marginal instability, classifying it as a saddle point.

$C_2$: This point corresponds to a potential dominated de Sitter phase, characterized by exponential expansion. Although it resembles a dark energy dominated universe, the existence of zero eigenvalues implies non-hyperbolic behavior, and the point remains a saddle. Hence, it cannot serve as a robust late-time attractor without additional stabilization mechanisms.

Across all three potentials, the LMG model admits: kinetic energy dominated early-time solutions, matter like scaling solutions, and scalar field dominated solutions. However, all critical points are saddles, indicating that the model lacks a natural late-time stable attractor when $\delta \neq 0$. This highlights the need for higher order Galileon terms to achieve a viable late-time cosmology.
\subsection{Quintessence: $\delta=0$}
\label{sec:d0}
\noindent The Galileon field action reduces to the standard quintessence in the limit $\delta=0$. In this case, the evolution of the Galileon field coincides with that of canonical quintessence throughout the cosmic history. Quintessence with $cosh$ and linear  potentials constitutes a well motivated dark energy scenario, as it naturally combines early-time scaling behavior with late-time accelerated expansion. This class of models provides a flexible framework for cosmic evolution, prevent fine-tuning issues, and remains compatible with cosmological observations. In this subsection, we consider the three previously introduced potentials for Quintessence, and analyze their stationary points along with the corresponding stability properties. Subsequently, we compare these results with those obtained for the LMG model. For $\delta=0$, the autonomous system (\ref{eq:auto}) reduces to a simplified set of evolution equations. The stationary points of the system are obtained by setting the left-hand side of equation (\ref{eq:auto}) to zero for each model. The stability of these critical points is subsequently determined by analyzing the signs of the eigenvalues of the corresponding Jacobian matrix. The resulting critical points along with their eigenvalues are summarized in Tables \ref{tab4},  \ref{tab5} and  \ref{tab6} for models (\ref{eq:Vm1}),  (\ref{eq:Vm2}) and  (\ref{eq:Vm3}), respectively.  Below is a systematic and detailed dynamical interpretation of all critical points appearing in Tables  \ref{tab4}- \ref{tab6}, corresponding to the Quintessence limit ($\delta=0$). The discussion highlights their physical nature, stability and cosmological relevance.

\begin{figure}
\begin{center}
\begin{tabular}{c c }
{\includegraphics[width=2.3in,height=2.1in,angle=0]{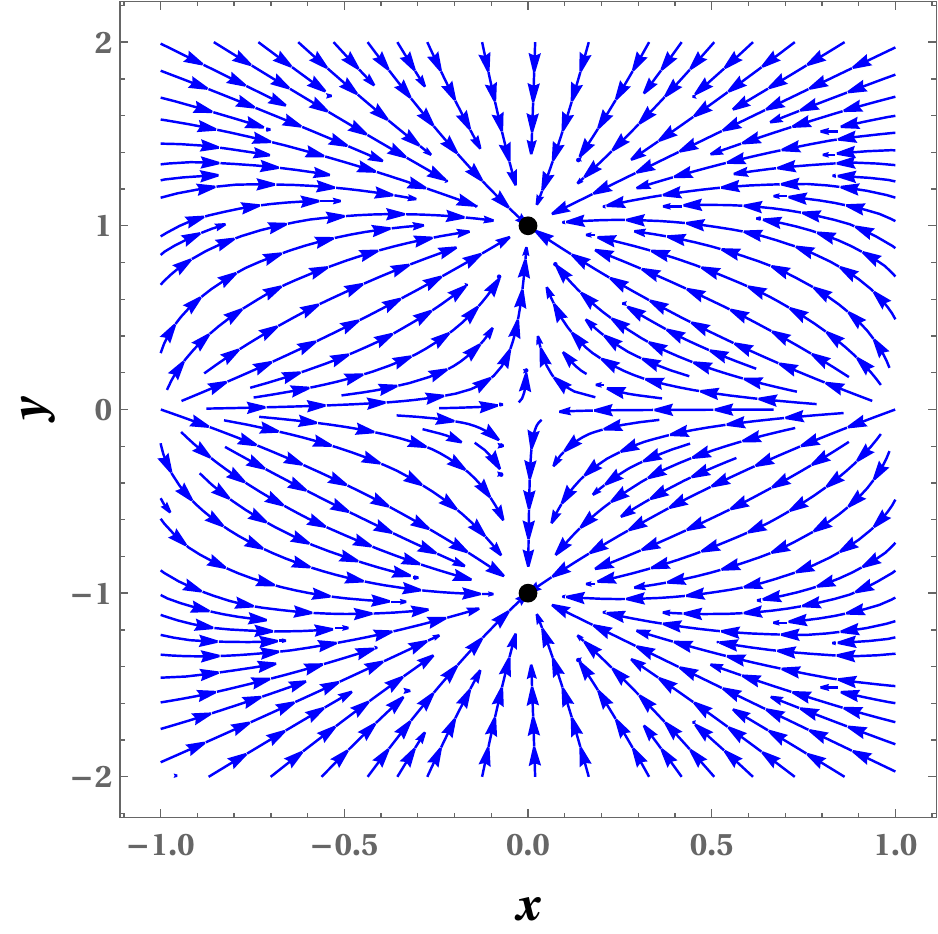}}&
{\includegraphics[width=2.3in,height=2.1in,angle=0]{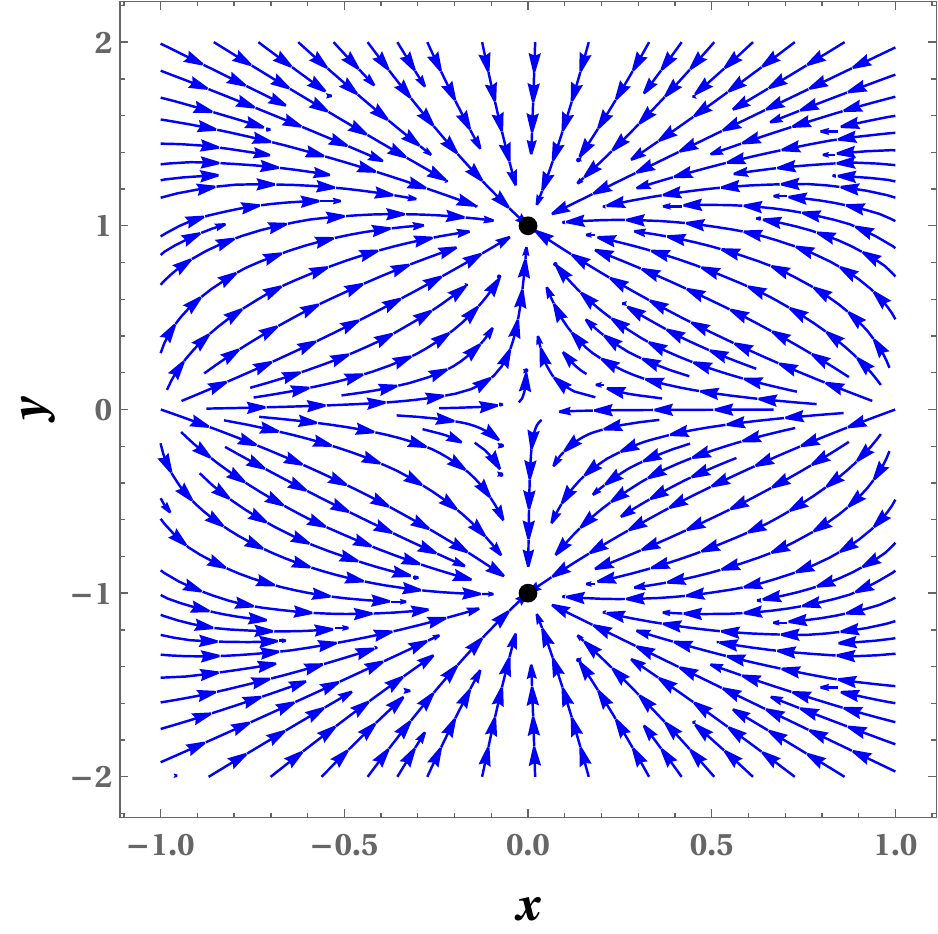}}
\end{tabular}
\caption{The figure illustrates the phase–space trajectories in the 
$(x,y)$ plane for points ($D_1$; left panel) and ($E_1$; right panel) within the framework of standard quintessence (see Eq. (\ref{eq:action}) with $\delta=0$, and subsection \ref{sec:d0}). The stable critical points act as attractive nodes, corresponding to $p=0.1$, $\alpha=1$, and $\beta=1$, respectively. The black dots indicate the stable attractor points toward which all trajectories converge.}
\label{fig:0}
\end{center}
\end{figure}

$D_1$: This point corresponds to a potential dominated scalar field configuration, where the kinetic energy vanishes. The universe undergoes exponential expansion, mimicking a cosmological constant. All eigenvalues are negative, ensuring linear stability. Hence, $D_1$ is a late-time attractor, capable of describing the observed accelerated expansion. The phase portrait for this point is shown in the left panel of Fig. \ref{fig:0}.

$D_2$: This represents a kinetic energy dominated solution, and behaves as a stiff fluid. The eigenvalues include positive and negative real parts, rendering the point a saddle node. It is therefore dynamically unstable and relevant only to early-time cosmology.

$D_3$: This point describes a matter dominated epoch, with the scalar field contributing negligibly to the total energy density. The presence of a zero eigenvalue makes the point non-hyperbolic, while the remaining eigenvalues indicate saddle behavior. It can therefore represent a transient matter era, but not a stable phase.

$D_4$: At this point, the scalar field energy density scales with the matter component, effectively behaving as pressureless dust. Such solutions are important for addressing the coincidence problem. However, since at least one eigenvalue is positive, the point is a saddle, describing only an intermediate phase of evolution.

$D_5$: This solution corresponds to a scalar field dominated universe, which may drive accelerated expansion for $p^2 \alpha^2 < 2$. Nonetheless, the mixed sign eigenvalues imply that the point is a saddle.

$E_1$: This point is the analog of $D_1$ and describes a stable dark energy dominated universe. All eigenvalues are negative, making it a late-time stable attractor. The phase portrait for this point is displayed in the right panel of Fig. \ref{fig:0}.

$E_2$: The Universe behaves as a stiff fluid dominated by scalar kinetic energy. The positive eigenvalues classify the point as a saddle, relevant only during early cosmic times.

$E_3$: This point corresponds to a standard matter era, with the scalar field frozen. The saddle nature allows it to represent a temporary phase during cosmological evolution.

$E_4$: This solution again represents matter like scaling behavior of the scalar field. Despite its phenomenological importance, the point remains a saddle.

$E_5$: This solution allows accelerated expansion for $\beta^2 < 2$, but due to its saddle nature it cannot act as a stable late-time attractor.

$F_1$: This stiff fluid solution is unstable, characterized by zero and positive eigenvalues, and hence classified as a saddle point.

$F_2$: This point corresponds to a potential dominated accelerating phase. However, the presence of a zero eigenvalue makes it non-hyperbolic, and it remains a saddle rather than a true attractor.

$F_3$: This point describes a standard matter epoch. The mixed eigenvalue structure confirms its saddle character, allowing it only as a transient cosmological phase.

In the Quintessence limit ($\delta=0$), the phase space exhibits: stable de-Sitter attractors for non-linear potentials ($D_1$ and $E_1$), early time kinetic dominated solutions ($D_2$, $E_2$ and $F_1$), transient matter and scaling solutions ($D_3$, $D_4$, $E_3$, $E_4$ and $F_3$), and saddle type scalar field dominated points ($D_5$, $E_5$ and $F_2$). Unlike the LMG case, Quintessence admits genuine late-time stable accelerating solutions, making it cosmologically viable within appropriate parameter ranges.
\begin{table}[tbp]
\caption{Displayed in the table are the stationary points, eigenvalues, and cosmological parameters corresponding to the Quintessence scenario (Eq. (\ref{eq:action}); $\delta = 0$) for potential (\ref{eq:Vm1}). Throughout the table, we take $\gamma_3=\sqrt{p^6 \alpha^6(24-7 p^2 \alpha^2)}$.}
\begin{center}
\begingroup
\setlength{\tabcolsep}{6pt} 
\renewcommand{\arraystretch}{1.5} 
\begin{tabular}{c c c c c c c c}
\hline
Point  & x  & y &  $\lambda$  & Eigenvalues & Stability  & $\Omega_{\phi}$  & $w_{eff}$\\
\hline
$D_1$ & 0 & $\pm 1$ & 0  & $
\begin{array}{l}
\mu_1 = -3 \\
\mu_2 = -\frac{3+\sqrt{9-12 p \alpha^2}}{2} \\
\mu_3 = -\frac{3-\sqrt{9-12 p \alpha^2}}{2} \\
\end{array}
$ & Stable & 1 & $-1$\\\\
$D_2$ & $\pm 1$ & 0 &  $\pm p \alpha$ & $
\begin{array}{l}
\mu_1 = 3 \\
\mu_2 = \sqrt{6} \alpha \\
\mu_3 = \frac{6-\sqrt{6} p \alpha}{2} \\
\end{array}
$ & Saddle & 1 & 1\\\\
$D_3$ & 0 & 0 &  $\pm p \alpha$ & $
\begin{array}{l}
\mu_1 = 0 \\
\mu_2 = -3/2 \\
\mu_3 = -3/2 \\
\end{array}
$ & Saddle & 0 & 0\\\\
$D_4$ & $\pm \sqrt{\frac{3}{2 p^2 \alpha^2}}$ & $\pm \sqrt{\frac{3}{2 p^2 \alpha^2}}$  & $\pm p \alpha$ & $
\begin{array}{l}
\mu_1 = 3/p \\
\mu_2 =-\frac{3(p^4 \alpha^4+\gamma_3)}{4 p^4 \alpha^4} \\
\mu_3 = -\frac{3(p^4 \alpha^4-\gamma_3)}{4 p^4 \alpha^4} \\
\end{array}
$ & Saddle & $\frac{3}{p^2 \alpha^2}$ & 0\\\\
$D_5$ & $\pm \frac{p \alpha}{\sqrt{6}}$ & $\pm \sqrt{1-\frac{p^2 \alpha^2}{6}}$ & $\pm p \alpha$ & $
\begin{array}{l}
\mu_1 = p \alpha^2 \\
\mu_2 =p^2 \alpha^2-3 \\
\mu_3 = \frac{p^2 \alpha^2-6}{2} \\
\end{array}
$ & Saddle & 1 & $\frac{p^2 \alpha^2}{3}-1$\\
\hline
\end{tabular}
\endgroup
\label{tab4}
\end{center}
\end{table}
\begin{table}[tbp]
\caption{The stationary points, eigenvalues, and cosmological parameters associated with the Quintessence (Eq. (\ref{eq:action}); $\delta = 0$) and potential (\ref{eq:Vm2}) are shown in the table, with $\gamma_4=\sqrt{\beta^6(24-7 \beta^2)}$.}
\begin{center}
\begingroup
\setlength{\tabcolsep}{6pt} 
\renewcommand{\arraystretch}{1.5} 
\begin{tabular}{c c c c c c c c}
\hline
Point  & x  & y & $\lambda$  & Eigenvalues & Stability  & $\Omega_{\phi}$  & $w_{eff}$\\
\hline
$E_1$ & 0 & $\pm 1$ & 0  & $
\begin{array}{l}
\mu_1 = -3 \\
\mu_2 = -\frac{3+\sqrt{9-12 \beta^2}}{2} \\
\mu_3 = -\frac{3-\sqrt{9-12 \beta^2}}{2} \\
\end{array}
$ & Stable & 1 & $-1$\\\\
$E_2$ & $\pm 1$ & 0 &  $\pm \beta$ & $
\begin{array}{l}
\mu_1 = 3 \\
\mu_2 = \sqrt{24} \beta \\
\mu_3 = \frac{6-\sqrt{6} \beta}{2} \\
\end{array}
$ & Saddle & 1 & 1\\\\
$E_3$ & 0 & 0 &  $\pm \beta$ & $
\begin{array}{l}
\mu_1 = 0 \\
\mu_2 = -3/2 \\
\mu_3 = -3/2 \\
\end{array}
$ & Saddle & 0 & 0\\\\
$E_4$ & $\pm \sqrt{\frac{3}{2 \beta^2}}$ & $\pm \sqrt{\frac{3}{2 \beta^2}}$ &  $\pm \beta$ & $
\begin{array}{l}
\mu_1 = 6 \\
\mu_2 =-\frac{3(\beta^4+\gamma_4)}{4 \beta^4} \\
\mu_3 = -\frac{3(\beta^4-\gamma_4)}{4 \beta^4} \\
\end{array}
$ & Saddle & $\frac{3}{\beta^2}$ & 0\\\\
$E_5$ & $\pm \frac{\beta}{\sqrt{6}}$ & $\pm \sqrt{1-\frac{\beta^2}{6}}$ &  $\pm \beta$ & $
\begin{array}{l}
\mu_1 = 2 \beta^2 \\
\mu_2 =\beta^2-3 \\
\mu_3 = \frac{\beta^2-6}{2} \\
\end{array}
$ & Saddle & 1 & $\frac{\beta^2}{3} -1$\\
\hline
\end{tabular}
\endgroup
\label{tab5}
\end{center}
\end{table}
\begin{table}[tbp]
\caption{The table summarizes the stationary points together with their eigenvalues and the corresponding cosmological parameters for Quintessence (Eq. (\ref{eq:action}); $\delta = 0$) in the presence of linear potential. (\ref{eq:Vm3}).}
\begin{center}
\begingroup
\setlength{\tabcolsep}{6pt} 
\renewcommand{\arraystretch}{1.5} 
\begin{tabular}{c c c c  c c c c}
\hline
Point  & x  & y &  $\lambda$  & Eigenvalues & Stability  & $\Omega_{\phi}$  & $w_{eff}$\\
\hline
$F_1$ & $\pm 1$ & 0 & 0 &  $
\begin{array}{l}
\mu_1 = 0 \\
\mu_2 = 3 \\
\mu_3 = 3 \\
\end{array}
$ & Saddle & 1 & 1\\\\
$F_2$ & 0 & $\pm 1$ & 0 &  $
\begin{array}{l}
\mu_1 = 0 \\
\mu_2 =-3 \\
\mu_3 = -3\\
\end{array}
$ & Saddle & 1 & $-1$\\\\
$F_3$ & 0 & 0 & 0 &  $
\begin{array}{l}
\mu_1 = 0 \\
\mu_2 =3/2 \\
\mu_3 = -3/2\\
\end{array}
$ & Saddle & 0 & 0\\
\hline
\end{tabular}
\endgroup
\label{tab6}
\end{center}
\end{table}
\section{Cosmological Implications, Results and Discussion}
\label{sec:CID}
\noindent In this work, we performed a comprehensive phase space analysis of the LMG scalar field model ($\delta \neq 0$) and its Quintessence counterpart ($\delta = 0$) for three different choices of scalar field potentials. The stationary points, their stability properties, and the associated cosmological parameters are summarized in Tables \ref{tab1}--\ref{tab6}. In this section, we discuss the main cosmological implications emerging from these results and highlight the key differences between the two scenarios.

For all considered models, we consistently obtain critical points corresponding to kinetic energy dominated scalar field configurations (e.g., points $A_1$, $B_1$, $C_1$, $D_2$, $E_2$ and $F_1$). These solutions are characterized by $\Omega_\phi = 1$ and $w_{\rm eff} = 1$, which corresponds to a stiff fluid dominated Universe. The eigenvalue analysis shows that these points are saddle nodes. Consequently, they are dynamically relevant only during the very early stages of cosmic evolution. This behavior is consistent with standard expectations in scalar field cosmology, where kinetic domination naturally precedes the radiation and matter dominated eras. For cosh type potentials, both the LMG and Quintessence models admit scaling solutions (e.g., points $A_2$, $B_2$, $D_4$ and $E_4$). These solutions describe a regime in which the scalar field energy density scales with the matter component, effectively mimicking cold dark matter. However, the corresponding eigenvalues indicate that these points are saddle points. As a result, they cannot act as late-time attractors. Scalar field dominated critical points ($A_3$, $B_3$, $D_5$ and $E_5$) appear in both the LMG and Quintessence frameworks, and are characterized by $\Omega_\phi = 1$ and $w_{\rm eff} = \frac{\lambda^2}{3} - 1$. Depending on the parameter values, these solutions may correspond to accelerated expansion when $\lambda^2 < 2$. In Quintessence case ($\delta = 0$), these points are typically saddle points, except for de-Sitter type solutions discussed below. In LMG scenario ($\delta \neq 0$), all scalar field dominated solutions also remain saddles, indicating that the Galileon coupling modifies the phase space structure but does not stabilize the late-time accelerated expansion.

A key distinction between the two models arises in the existence and stability of de-Sitter solutions. In Quintessence framework, critical points $D_1$ and $E_1$ satisfy $w_{\rm eff} = -1$ and $\Omega_\phi = 1$, and are found to be stable attractors. These solutions naturally describe a dark energy dominated Universe in agreement with current observational data. In contrast, within the LMG model, although de-Sitter like points (e.g., $C_2$) do exist, they are non-hyperbolic saddle points. Consequently, they cannot serve as robust late-time attractors. This highlights an important cosmological implication: while standard Quintessence admits a stable accelerating solution, the LMG model does not provide a natural late-time attractor for all the potentials considered. For the linear potential, both LMG and Quintessence scenarios yield only saddle critical points, as shown in Tables \ref{tab3} and \ref{tab6}. Although de-Sitter configurations appear, they remain dynamically unstable. This result indicates that linear potential alone is insufficient to support a stable accelerating phase, irrespective of the presence of Galileon correction term.

Galileon models exhibit a rich phenomenology in theoretical physics arising from their extended parameter space. The Refs. \cite{br1,br1a} indicate that certain regions of the parameter space are already excluded by observational data. Barreira et al. demonstrated that Galileon models exhibit serious tensions with Integrated Sachs–Wolfe (ISW) observations. More recently, a comprehensive analyses combining Cosmic Microwave Background (CMB), Baryon Acoustic Oscillation (BAO), and ISW data have placed stringent constraints on Cubic, Quartic, and Quintic Galileon cosmologies. These studies rule out the Cubic Galileon model and emphasize the important role of neutrino mass in alleviating observational tensions \cite{br2,br3,br4}. For further details, readers may refer to Refs. \cite{br5,br6,br7}. 

In our analysis, we find that the LMG model does not admit a stable late-time attractor solution. Furthermore, perturbative constraints from ISW effect reinforce the conclusion that the LMG framework is not phenomenologically viable, unless higher order Galileon terms are incorporated. In the shift symmetric scenario, cosh-type potentials explicitly break the Galileon symmetry, whereas a linear potential preserves it in flat spacetime. However, within the context of our study, the inclusion of a potential term does not change the results. Recently, Tsujikawa demonstrated an explicit model with broken shift symmetry, where the scalar field includes a potential alongside a Galileon self-interaction and a quadratic kinetic term. This framework successfully realizes the desired phantom-divide crossing at low redshifts while avoiding ghost and Laplacian instabilities \cite{st1}. In LMG scenario, the absence of stable accelerating attractors prevents a comparison with observational constraints, which further limits the phenomenological viability of the model. In contrast, within the Quintessence framework, the critical points $D_1$ and $E_1$ show de-Sitter features. Though, the EoS for both points give rise to $-1$, indicating that they do not display a phantom character.
\section{Conclusion}
\label{sec:conc}
\noindent In this work, we restrict our analysis to the lowest order Galileon Lagrangian and supplement it with a general scalar potential. We have carried out a detailed dynamical system analysis of the LMG model $(\delta \neq 0)$ and its Quintessence counterpart $(\delta = 0)$, with the aim of assessing their viability as candidates for late--time cosmic acceleration. Restricting attention to a spatially flat FLRW universe, we analyzed the cosmological dynamics arising from the cubic Galileon interaction supplemented by three different scalar field potentials, and compared the results with those obtained in the standard Quintessence. Our analysis reveals that the LMG framework admits critical points corresponding to kinetic energy dominated regimes, matter like scaling solutions and scalar field dominated configurations. However, the linear stability analysis shows that all such points are saddle  in phase space. As a consequence, the LMG model does not possess a natural late--time stable attractor capable of describing the observed accelerated expansion of the Universe. Even though de-Sitter like solutions appear for certain potentials, their non-hyperbolic nature prevents them from acting as robust attractors, see Tables \ref{tab1}, \ref{tab2} and \ref{tab3}.

In contrast, the Quintessence scenario exhibits a richer and more phenomenologically viable phase space structure. For non-linear potentials, we identify stable de-Sitter attractors characterized by $\Omega_\phi = 1$ and $w_{\rm eff} = -1$, which can successfully account for dark energy dominated epoch, see points $D_1$ and $E_1$. The phase portraits shown in Fig. \ref{fig:0} further illustrate that all trajectories converge toward the attractor points, confirming their stability and relevance for late--time cosmic evolution. This qualitative difference highlights the crucial role played by the Galileon coupling in modifying the asymptotic structure of the phase space, often at the expense of late--time stability. These critical points are summarized in Tables \ref{tab4}, \ref{tab5} and \ref{tab6}.

Overall, our results indicate that while the LMG model significantly alters cosmological dynamics and phase space behavior, it fails to provide a stable late--time accelerating solution for the class of potentials studied here. This suggests that either more general scalar potentials or the inclusion of higher-order Galileon/Horndeski interactions may be necessary to reconcile Galileon cosmology with observational requirements. We expect that such extensions could shed further light on the viability of modified gravity scenarios as alternatives to standard dark energy models.

\section*{Acknowledgments}
\noindent MS acknowledges Integral University, Lucknow for financial support through Seed Money Grant 2024-2025 (Project Sanction No.: IUL/ICEIR/SMP/2024-04) and MCN: IU/R\&D/2026-MCN0004329. MS also thanks the Inter-University Centre for Astronomy and Astrophysics (IUCAA), Pune for the hospitality and facilities under the visiting associateship program where the work was initiated. 

\end{document}